\documentclass[%
 aip,
 amsmath,amssymb,
 reprint,%
graphicx
]{revtex4-1}

\usepackage{graphicx}% Include figure files
\usepackage{dcolumn}% Align table columns on decimal point
\usepackage{bm}% bold math
\usepackage[utf8]{inputenc}
\usepackage[T1]{fontenc}
\usepackage{mathptmx}

\usepackage{xfrac}
\usepackage{multirow}
\usepackage[table,svgnames,xcdraw]{xcolor}
\usepackage{colortbl}
\usepackage{xcolor}
\usepackage{tikz}
\usepackage{subfigure}
\usepackage[colorlinks=true,linkcolor=cyan]{hyperref}%
\begin{document}

\preprint{AIP/123-QED}

\title[]{New Way of Generating Electromagnetic Waves }
% Force line breaks with \\

\author{Ali Hossseini-Fahraji}
\author{Majid Manteghi}
\author{Khai D. T. Ngo}
\affiliation{ 
Bradley Department of Electrical and Computer Engineering, Virginia Tech, Blacksburg, Virginia 24061, USA
%\\This line break forced with \textbackslash\textbackslash
}%

\date{\today}% It is always \today, today,

\begin{abstract}
This paper presents a new method for generating low-frequency electromagnetic waves for navigation and communication in challenging environments, such as underwater and underground. The main idea is to store magnetic energy in two different spaces using the interaction between a permanent magnet and a magnetic material. The magnetic reluctance of the medium around the permanent magnet is modulated to change the magnetic flux path. The nonlinear properties of magnetic material as a critical phenomenon are used for effective modulation. As a result, a time-variant field is generated by the modulation of the permanent magnet flux. This non-resonant time-variant characterization means that the transmitter is not bound to the fundamental limits of the antennas and can transmit higher data rates. A prototype transmitter as a prove-of-concept is designed and tested based on the proposed idea. Compared to the rotating magnet, the prototyped transmitter can modulate $50\%$ of the stored energy of the permanent magnet with much lower power consumption.
\end{abstract}

\maketitle

\section{\label{sec:introduction} Introduction/Background}
\vspace{-1em}
The emphasis is primarily on increasing data rate, which leads to the use of higher frequencies and wider bandwidths in modern communication technology research and innovations. However, because of physical constraints, increasing frequency and bandwidth in many areas of technology cannot necessarily be beneficial. Communication under seawater or other challenging RF environment require very low-frequency, VLF, or ultra-low-frequency, ULF signals to penetrate lossy media that block high-frequency signals. Also, new developments in neuroscience have shown the potentials of ULF and VLF electromagnetic, EM, waves to treat neurological conditions such as Alzheimer's disease, amyotrophic lateral sclerosis, persistent vegetative diseases, epilepsy, stroke-related illness, tinnitus, multiple sclerosis, schizophrenia, and traumatic brain injury. The main challenge is that most of VLF and ULF generators are large and power-hungry, which make them impractical or hard to use in many applications. In this paper, we present a new approach for generating EM waves in a compact and low-power fashion.

At first, radio wave technology was developed within the VLF ranges. Because of the broad spectrum of radiated waves and the problem of spark-gap oscillators (invented by Hertz) interference, William Crookes proposed using sinusoidal sources in resonance structures (then called syntony) to minimize the transmitter and receiver bandwidth in 1892 \cite{Crookes92}. It started a race to develop a continuous wave, CW, sinusoidal wave generator to replace the spark-gap sources for RF applications. Innovative structures were proposed by several researchers (Elihu Thomson, Nikola Tesla, Reginald Fessenden, and many others). Finally, the spark-gap oscillators were replaced by the Alexanderson alternator (a mechanical structure based on a rotating permanent magnet) in 1904. Surprisingly, many variants of Alexanderson alternator have been suggested after more than a century \cite{Fawole17,Strachen18,Burch18,Gong18,Golk18,Prasad18,Bickford19}, in response to a DARPA call for ELF and VLF sources in recent years. Such mechanical generators (mechtenna), however, still have the same shortcomings as the original design, such as large size, massive power consumption, hard to modulate and transmit information, synchronization, noise, vibration, and durability problem of a mechanical structure. There have also been other versions of mechanical vibration proposed to generate EM waves in VLF ranges as well \cite{Trouton92,Datskos12,Madan17,Wang18,Scott19}.

In 1961, on the other hand, an analysis of EM radiation from the acoustically-driven ferromagnetic yttrium iron garnet sphere (YIG) introduced the concept of acoustic resonance as strain powered (SP) antenna. Recent studies have shown that in a device with smaller physical dimensions than the EM wavelength, multiferroic antennas can take advantage of acoustic resonance to reduce antenna size \cite{Nan08,Nan17,Domann17}. As a contrast to rotating permanent magnets, strain-coupled piezoelectric and magnetostrictive composites are thus used in magnetostrictive materials to control magnetic spin states \cite{Ueno03,Xu19,Schneider19}. Although this technique removes the necessary inertial force in mechtenna, it faces challenges due to the matching rigidity between piezoelectric and magnetostrictive (i.e., low energy transfer from piezoelectric to magnetostrictive) and sequentially inefficient power transfer to electromagnetic radiation. Besides, making this structure into bulk is also a challenge. As an alternative technique, we intend to use magnetic material to manipulate the magnetic flux of a permanent magnet.

This idea is to alter the reluctance of the flux path to make the magnetic flux time-variant by pushing it to take an alternative path. Our concept is based on `variable material' rather than `variable structure' as in mechanical rotation. We take advantage of a permanent magnet, which is equivalent to a lossless electromagnet with the winding of the superconductor, which produces a static magnetic flux without dissipating power. Meanwhile, we alternate the direction of flux between free space and a medium with high permeability. The permeability of the magnetic material near the permanent magnet varies by adjusting the current through a control coil, depending on the B-H curve of the magnetic material \cite{Strachen18}.

There are many papers published in the last three years on ULF antennas; however, most of them have not evaluated their work with a concrete criterion. Therefore the performances of these proposed antennas are difficult to assess and compare. We consider a permanent magnet's magnetic flux to be a suitable reference to evaluate the performance of any ULF transmitter. Hence, from now on, we believe the field produced by a rotating magnet to be a reference to assess the field generated by any technique with the same volume magnet. In this way, we calibrate the receiving device (searching coil or any other type of magnetometer), especially if we can rotate the magnet to the generator's operating frequency. Also, we suggest calculating the leakage of the windings around the ferrite cores independently of the permanent magnet to be able to distinguish between the permanent magnet's contributions and the entire field of windings.

In this research, the magnetic flux per volume of selected published designs is compared in \textcolor{cyan}{Table~\ref{tab:table1}} to give a better estimate of the performance of our design. Note that most articles do not provide details about the antenna's total volume, and the information is limited to the size of the main radiating element. The objective of this comparison is to determine the minimum volume needed to reach a field strength of 1 fT at 1 km.

As shown in \textcolor{cyan}{Table~\ref{tab:table1}}, the results for the radiator volume of $1\,cm^3$ ($\Delta$B/Vrad) show that the rotating magnet has the maximum magnetic flux, as expected. Without any modulation, the rotating magnet generates a magnetic flux of about $200 \times 10^{-3}\, fT/cm^3$, whereas any designs aimed at modulating magnet rotation reduced its efficiency significantly. Furthermore, our proposed design and the best multiferroic antenna design in the literature can generate $98 \times 10^{-3}$ and $13.3 \times 10^{-3}\, fT/cm^3$ magnetic fluxes corresponding to 49\% and 13\% efficiency of these antennas, respectively. The results show at the time of publication that the proposed design has the best chance to compete with a rotating magnet with considerably lower power consumption and smaller size.
\vspace{-1em}

\begin{table*}
\small
\caption{\label{tab:table1} Comparison of different low-frequency antennas for use in  underwate and underground communication.}
\resizebox{0.97\textwidth}{!}{%
% \squeezetable
\begin{ruledtabular}

\begin{tabular}{l|l|l|l|l|l|l|l|l}
% 1st row
{\bf Method} &  {\bf Ref} & {\bf Modulation \footnote{EMR: Electrically Modulated Reluctance, DAM: Direct Antenna Modulation}} & \begin{tabular}[c]{@{}l@{}}{\bm{ $V_{rad}$}\footnote{The volume of the central radiator}}\\{\bf $\left (cm^3 \right)$}\end{tabular} & \begin{tabular}[c]{@{}l@{}}{\bm{$V_{Antenna}/ D$} \footnote{This column describes the total size of the antenna and the largest antenna dimension extracted from the literature, where applicable.}}\\{\bf $\left (cm^3 / cm \right)$}\end{tabular} & {\bf Freq (Hz)} & {\bf $\Delta B$} & \begin{tabular}[c]{@{}l@{}}{\bf $\Delta B (fT)$}\\{\bf $at 1 km$}\end{tabular} & \begin{tabular}[c]{@{}l@{}}{\bf $\Delta B/ V_{rad}$ }\\{$\left (fT/ cm^3 \right) \times 10^{-3}$}\\{\bf $at 1 km$}\end{tabular}  \\ 
 \hline
% 2nd row
{} & ${}_{\cite{Gong18}}$ & - & \textsc{100} & -/10 & 30 & 1 pT at 264.8 m & 18.5 & 185.00  \\ \cline{2-9}
% 3rd row
{} & ${}_{\cite{Scott19}}$ & - & $100$ & - & 100 & 1800 nT at 2.03 m & 15.06 & 150.60 \\ \cline{2-9}
% 4th row
\multirow{-3}{*}{\begin{tabular}[c]{@{}l@{}}{\bf Rotating}\\ {\bf Magnet}\end{tabular}} & ${}_{\cite{Burch18}}$ & - & 3 & 3/1.6 & 500 & 800 fT at 100 m & 0.8000 & 266.67 \footnote{This value shows a higher value than the theory, which may be due to the magnetometer's error.} \\ \hline
% 5th row
{} & ${}_{\cite{Fawole17}}$ & Electromechanically & 8.4 & 58.5/15.6 & 22 & 600 nT at 1 m & 0.6000 & 71.43 \\ \cline{2-9}
% 6th row
{} & ${}_{\cite{Strachen18}}$ & EMR & 3.62 & 353/8 & 150 & 100 nT at 0.3 m & 0.0027 & 0.75 \\ \cline{2-9}
% 7th row
\multirow{-3}{*}{\begin{tabular}[c]{@{}l@{}}{\bf Modulated}\\ {\bf Magnet}\\ {\bf Rotation}\end{tabular}} & ${}_{\cite{Golk18}}$ & Mechanical Shutter & - & - & 960 & 1.3 nT at 1 m & 0.0013 & - \\ \hline
% 8th row
{\bf Pendulum Array} & ${}_{\cite{Prasad19}}$ & DAM & 29.91 & -/13.4 & 1030 & 79.4 fT at 20 m & 0.0006 & 0.02\\  \hline
% 9th row
{\bf Piezoelectric} & ${}_{\cite{Kemp19}}$ & DAM & 18.9 & -/9.4 & 35500 & - & - & - \\ \hline
% 10th row
{} & ${}_{\cite{Xu19}}$ & DAM & 6.4 & -/25 & 28000 & 16 nT at 0.4 m & 0.0010 & 0.16\\  \cline{2-9}
% 11th row
\multirow{-2}{*}{\bf Multiferroic} & ${}_{\cite{Schneider19}}$ & DAM & 1 & -/18 & 10 & 6.05 nT at 1.3 m & 0.0133 & 13.30 \\ \hline
% 12th row
{\bf Motionless} & - & EMR \& DAM & 3 & 280/14 & 430 & 170 nT at 1.2 m & 0.2940 & 98 

\end{tabular}
\end{ruledtabular}}
\end{table*}

\section{\label{sec:theory} Theoretical Background}
\vspace{-1em}
The traditional way of generating electromagnetic waves is to periodically exchange electric and magnetic energy stored in two distinct parts of the radiating system. Any or both types of stored energy may leak some power as radiation. We thus have a specific amount of radiated power, $P_{r}$, for a maximum amount of stored energy, $W_{max}$, and we can calculate the antenna's quality factor as $Q=\omega W_{max}/ P_{r}$. Fundamental limits of antennas \cite{Chu48,Wheeler47} tie an antenna's quality factor to its electrical size as $Q=1⁄ (ka)^3$, where \textit{a} is the radius of smallest surrounding sphere and $k=2\pi⁄\lambda$ is the wave number. That means the smaller the antenna, the more energy we need for a given radiated power to be stored. Moreover, the quality factor is related to the antenna's instantaneous bandwidth. The simple conclusion shows that we need to store a large amount of energy in the antenna reactive-zone in cases of low frequency or small antennas ($a⁄\lambda \ll 1$), and the instantaneous bandwidth will be small.

Instead of exchanging energy between electric and magnetic forms, a static stored energy (e.g., stored energy in a permanent magnet or an electret) can be moved, vibrated, or rotated without altering its form to generate a time-varying field. This approach differs radically from the traditional radiation systems and is therefore not constrained by resonance limitations. However, it may not be desirable to apply any of these approaches to magnets or electrets by using mechanical movements. We propose to modulate the magnetic energy stored around a magnet by manipulating reluctance to the surroundings. Therefore, the direction of the flux or the position of the stored energy variates in time. The magnetic field thus varies in time.

Let us first look at the magnetic flux density of a uniformly magnetized sphere, as shown in \textcolor{cyan}{Fig.~\ref{fig:fig1}}:

\begin{figure}
\includegraphics[clip, trim=2cm 21.3cm 15cm 2cm,width=0.45\columnwidth]{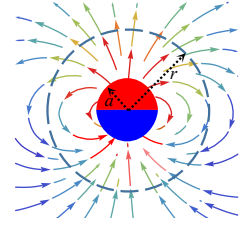}
\caption{\label{fig:fig1} Magnetic flux density decays by $\frac{1}{r^3}$.}
\end{figure}

\begin{equation}
B \left (r>a \right)=\frac{\mu_{0}}{4\pi}\left [\frac{-\bf{m}}{r^3}+\frac{3\left(\bf{m}.\bf{r} \right) \bf{r}}{r^5} \right] , \qquad {\bf m}=\frac{4}{3}\pi a^3 \bf{M}
\label{eq:one}
\end{equation}

where {\bf M} (A/m) is the magnetic dipole moment per unit volume of the permanent magnet. As is evident from the closed-form magnetic flux density of spherical magnet, there is a $1⁄ r^3$  decay for $r>a$. One can compute the total magnetic energy stored around the magnet as:

\begin{equation}
W_{m} = \frac{\mu_{0} \left |{\bf m} \right|^2}{12\pi a^3} = \frac{\mu_{0}}{9} V \left |{\bf M} \right|^2
\label{eq:two}
\end{equation}

where $V$ is the volume of the magnet with magnetization {\bf M}. One can compute the total energy stored outside a sphere of radius $r>a$ as: 

\begin{equation}
W_{r} = \left (\frac{a}{r} \right)^3 W_{m}
\label{eq:three}
\end{equation}

The above equations indicate that the magnetic energy contained in the radius $r$ sphere and the magnetic flux intensity in the distance $r$ decrease by $1/r^3$. Thus, in order to reduce the size of the transmitter, a high-magnetic flux (requires more sophisticated material) must be modulated when selecting a small $r$. Otherwise, miniaturization must be sacrificed in order to modulate smaller magnetic fluxes at larger $r$.

The first approach is to use a material with a controllable reluctance to create a shielding layer at radius \textit{r}. Ideally, one can alter the shield's reluctance from a small to substantial value. This process allows the stored energy to be temporarily decoupled outside the shield from the magnet, and then allows the magnet to store energy outside the shield again by increasing its reluctance. In its low reluctance mode, the spherical shield closes the field lines that pass it and thus dissipates the $W_{r}$ energy every half cycle. For analytical convenience, we presume that variation of the reluctance does not substantially disrupt the magnetic flux within the shield. There are various constraints, including loss, size, the current required to control the shielding material's reluctance, and saturation level, which dictate the proper values for \textit{r}.

\begin{figure}[t]
  \begin{center}
        \begin{tabular}{ m{0.4\columnwidth} m{0.5\columnwidth}  }
            \quad \raisebox{-.5\height}{\includegraphics[clip, trim=12cm 7.5cm 12.5cm 7.5cm,width=0.25\columnwidth]{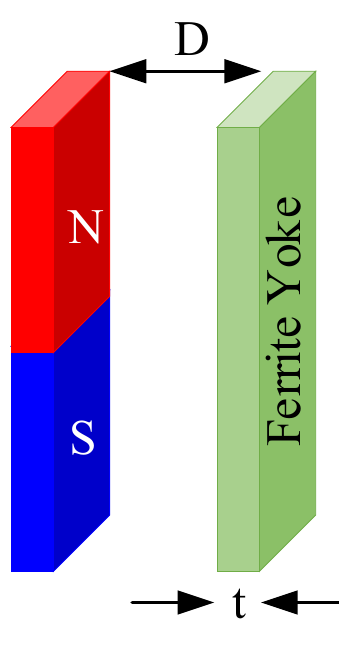}}
            & 
            \quad \raisebox{-.5\height}{\includegraphics[clip, trim=10.7cm 7cm 11cm 7cm,width=0.35\columnwidth]{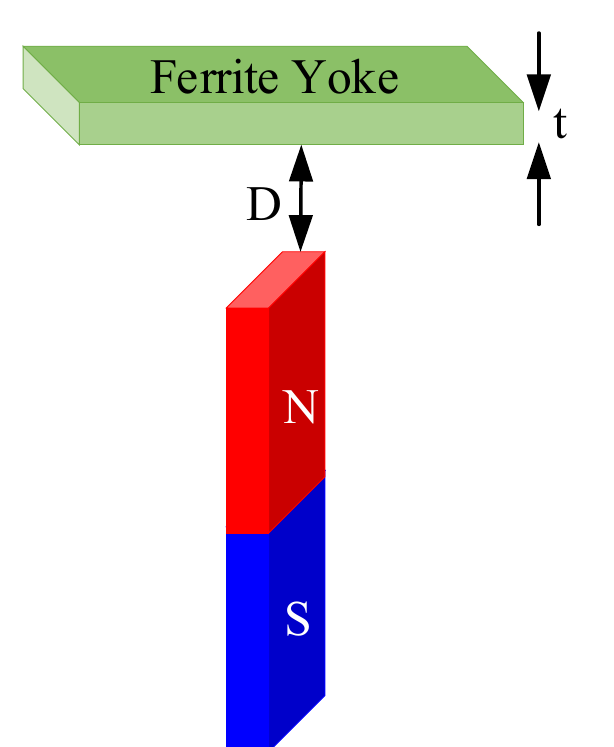}} \\
            \qquad \qquad (a) & \qquad \qquad (b) \\
            \multicolumn{2}{c}{\includegraphics[clip, trim=3.9cm 22.6cm 11.6cm 1.8cm,width=0.9\columnwidth]{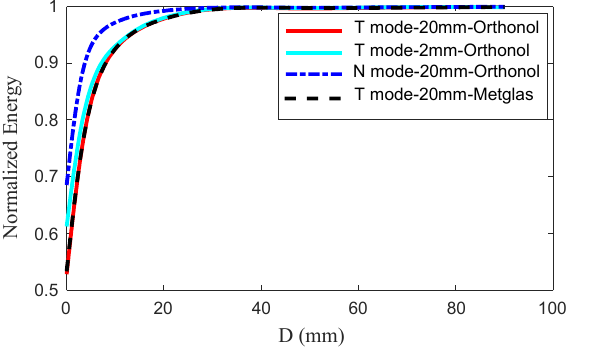}}\\
            \multicolumn{2}{c}{(c)}
        \end{tabular}
      \caption{The system contains magnet and ferrite yoke. \textit{a}) T mode \textit{b)} N mode \textit{c)} normalized system energy of the system compared to the isolated magnet versus distance D; the legend shows the system‘s mode, the thickness of the ferrite yoke (t), and the type of magnetic material used for ferrite yoke implementation.}
      \label{fig:fig2}
 \end{center}
\end{figure}

We consider the next approach to be an asymmetric system consisting of a ferrite yoke (as the variable reluctance magnetic material) and a permanent magnet (as the magnetic flux source), as shown in \textcolor{cyan}{Fig.~\ref{fig:fig2}(a)} and\textcolor{cyan}{~\ref{fig:fig2}(b)}. Since the permanent magnet attracts the ferrite yoke, the total energy stored in this system is a function of the distance from the yoke to the magnet. We simulated this structure using ANSYS Maxwell for different materials and ranges and compared the energy of the system with the energy stored in the isolated magnet. The simulation results, as shown in \textcolor{cyan}{Fig.~\ref{fig:fig2}(c)}, suggests that nearly half of the magnet's energy is converted to kinetic energy when the ferrite yoke contacts the magnet, and another half is still stored around the system. The system energy for $D = 1 cm$ is about 90\% of its maximum value, as the simulation results show. One can then move the yoke 1 cm away from the magnet back and forth and modulate the stored energy with a modulation depth of 40\%. We can use a mechanical resonance structure (i.e., a spring and a fixture) to conserve the kinetic energy. We intend to modulate the reluctance to make the stored energy time-variant, rather than a mechanical movement.

While the spherical shield offers a significant modulation depth (close to 100\%), it is large and three-dimensional. On the other hand, the system with ferrite yoke has much smaller dimensions; however, it cannot provide a sufficiently broad modulation depth. Therefore, we combine the two above methods by putting the magnet on a ferromagnetic film with a proper winding to modulate the magnetic flux by adjusting the reluctance of the film. The design parameters include the ferrite characteristics, in particular, the nonlinearity of its B-H curve, the thickness of the ferromagnetic film, the topology of the structure, and the windings. The design objectives are high magnetic flux, a high modulation depth, small size, and low dissipated power. One of the tasks to achieve these objectives is to utilize the relationship between magnetic flux, {\it {\bf B}}, and magnetic field, {\it {\bf H}}, effectively.
\vspace{-1em}

\section{\label{sec:design} Transmitter Design}
\begin{figure}
    \centering
    \subfigure[]
    {
        \includegraphics[clip, trim=2.5cm 21.8cm 8.5cm 2.8cm,width=0.9\columnwidth]{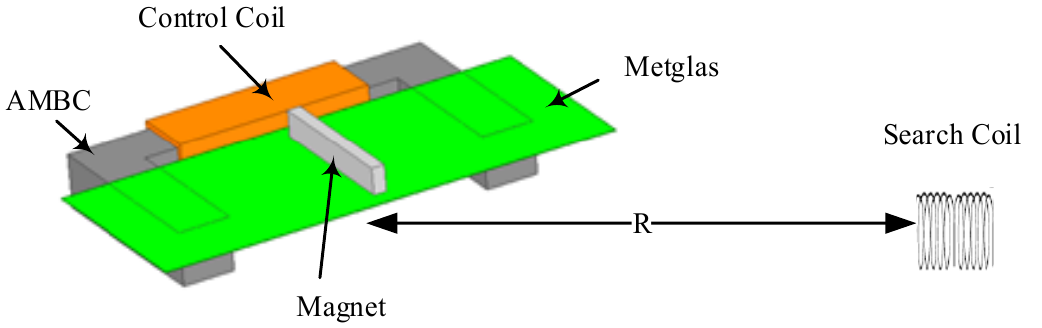}
        \label{fig:fig3(a)}
    }
    \\
    \subfigure[]
    {
        \includegraphics[clip, trim=2.5cm 21.8cm 14cm 2.5cm,width=0.9\columnwidth]{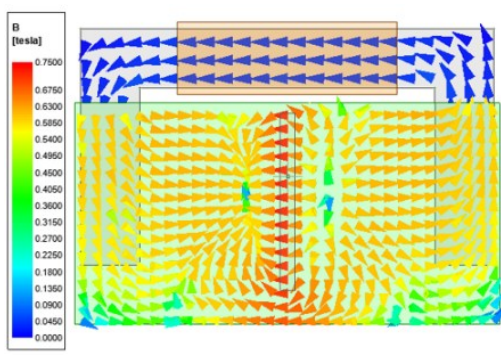}
        \label{fig:fig3(b)}
    }
    \subfigure[]
    {
        \includegraphics[clip, trim=2.5cm 14.5cm 3cm 2.8cm,width=0.9\columnwidth]{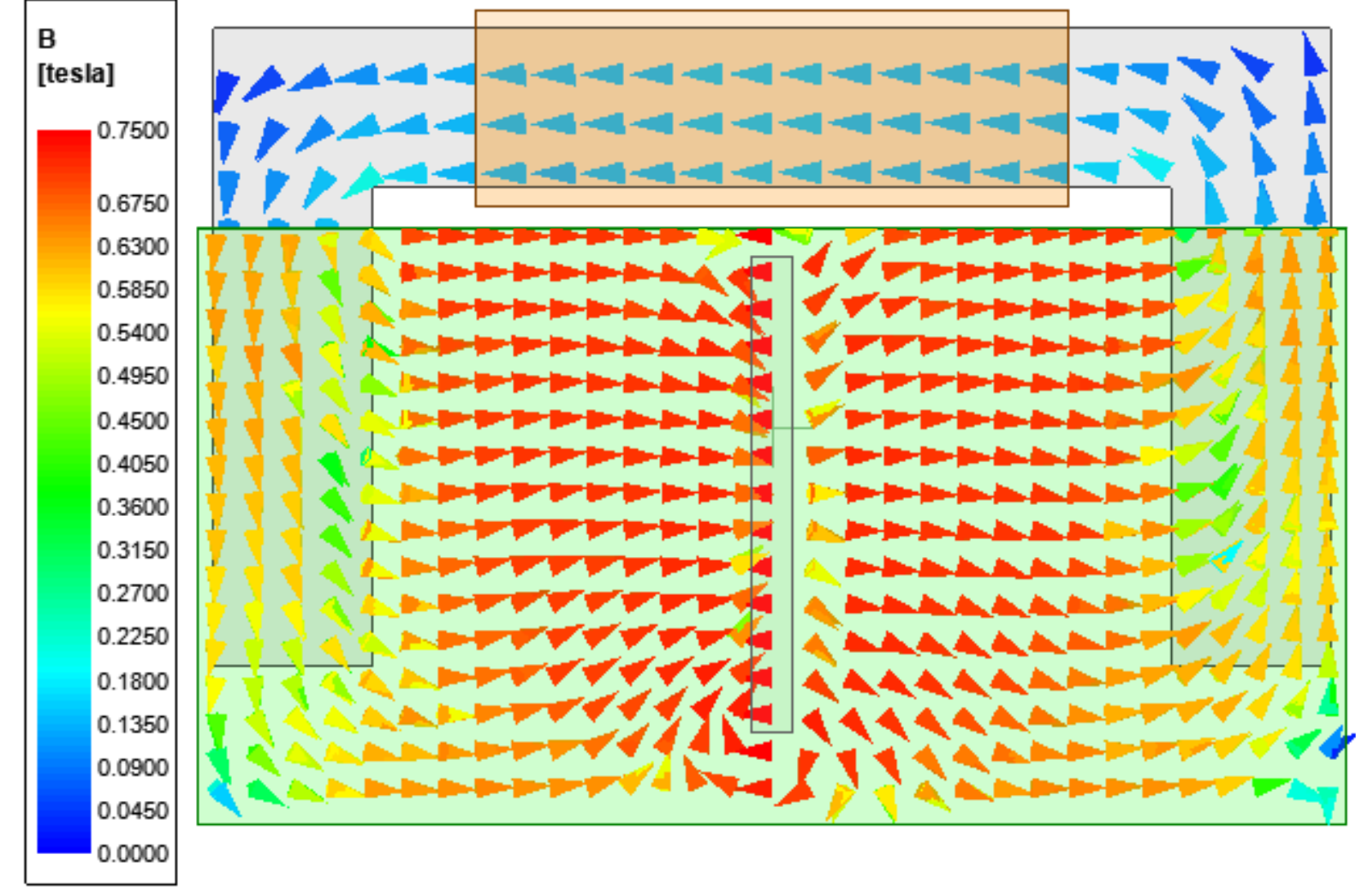}
        \label{fig:fig3(c)}
    }
\caption{ \textit{a}) Maxwell model of the prototyped system. Magnetic field distribution of the Metglas \textit{b}) control current is zero and the system is in closed mode \textit{c}) Control current force the Metglas to saturation (Open mode).}
\label{fig:fig3} 
\end{figure}

\vspace{-1em}
We designed and prototyped different structures to examine our proposed approach. \textcolor{cyan}{Figure~\ref{fig:fig3(a)}} shows the ANSYS model of one of our designs. The permanent magnet used in this transmitter is a rare-earth Neodymium magnet (N52, $6 \times 1 \times 0.5\, cm$), which is the strongest permanent magnet available in the market. Also, we used seven layers of Metglas sheets 2705M ($B_{s} = 0.77\,T$) with a total thickness of 0.178 mm as the magnetic film. Besides, a 40-turn coil around a c-shape magnetic core made of amorphous AMBC ($B_{s} = 1.56\,T$) with a $2 \times 2\,cm$ cross-section generates the magnetic flux needed to modulate the magnetic film's reluctance. We select a low reluctance core with a reasonably broad cross-section to ensure that the c-shape core works at its linear state. As a result, the current through the control coil generates a magnetic flux in the magnetic film.

\textcolor{cyan}{Figure~\ref{fig:fig3}} shows the flux density on the system for two different values for the control current. \textcolor{cyan}{Figure~\ref{fig:fig3(a)}} shows that small areas of the magnetic film are in saturation when the control current is zero. The saturated film helps to spread magnetic flux in the air and to store magnetic energy around the magnet. The small saturated area shows that the magnetic film operates as a barrier and closes inside the magnetic flux. Next, we apply 0.5 A current to the control coil, and the pattern of magnetic film saturation shifts to \textcolor{cyan}{Fig.\ref{fig:fig3(b)}}, which means the saturated area is larger than the closed mode. In this mode, the magnetic flux spreads more in space, and there is more energy stored around the magnet. We name this state of the system,``Open mode.'' This system's operating modes will differ by adjusting the arrangement of the magnet or the magnetic film. For example, the magnetic film may be saturated by a giant magnet with zero current. The saturated area of the magnetic film can then be reduced by a magnetic flux generated by the control current against the magnet's magnetic flux. In this case, the system's operating modes switch to open and closed mode for zero current and high current, respectively. One can apply a sinusoidal current to the control coil to change the amount of energy stored around the magnet periodically. \textcolor{cyan}{Figure~\ref{fig:fig3(b)}} also shows that the magnetic flux density in the AMBC core is less than 0.13 T, indicating the amorphous AMBC cross-section we have is higher than what we needed to keep it out of saturation. One can use a smaller core to reduce overall system size and weight.
\vspace{-1em}

\section{\label{sec:measurement} Measured Results}
\begin{figure}
  \centering
        \subfigure[]
    {
        \includegraphics[clip, trim=2.3cm 20.5cm 10.5cm 2.5cm,width=0.9\columnwidth]{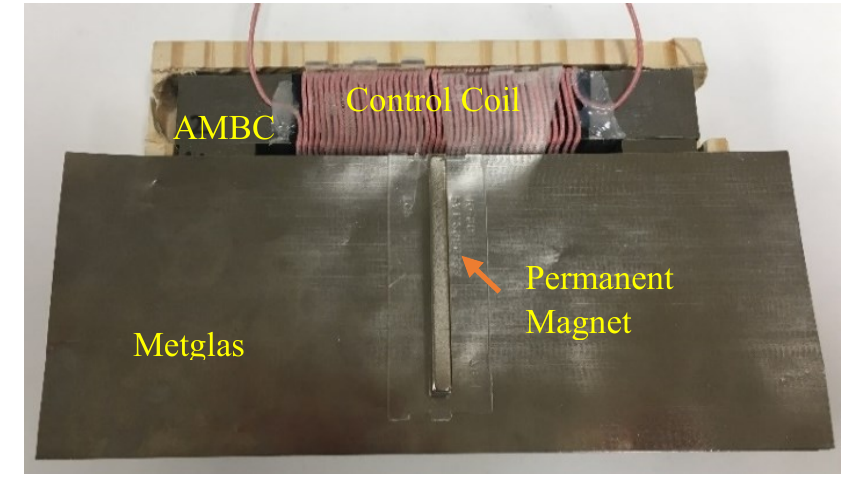}
        \label{fig:fig4(a))}
    }
        \subfigure[]
    {

        % \setbox1=\hbox{\includegraphics[clip, trim=3cm 20.5cm 10.5cm 2.5cm,width=0.9\columnwidth,height=4.5cm]{Fig/Fig4(b).pdf}}

        % \includegraphics[clip, trim=3cm 20.5cm 10.5cm 2.5cm,width=0.9\columnwidth,height=4.5cm]{Fig/Fig4(b).pdf}
        % % \llap{\makebox[\wd1][l]{\raisebox{1cm}{\includegraphics[clip, trim=3cm 24cm 12.5cm 2.5cm,height=1cm]{Fig/Fig4(c).pdf}}}}
        % \llap{\makebox[\wd1][l]{\raisebox{3cm}{\includegraphics[clip, trim=3cm 24cm 12.5cm 2.5cm,height=1.4cm]{Fig/Fig4(c).pdf}}}}
        %         \begin{tikzpicture}
        %          \draw (-2,-2) -- (-3,-3);
        %         \end{tikzpicture}
        % \label{fig:fig4(b)}
        \includegraphics[clip, trim=2.5cm 20.5cm 10.5cm 2cm,width=0.9\columnwidth,height=4.5cm]{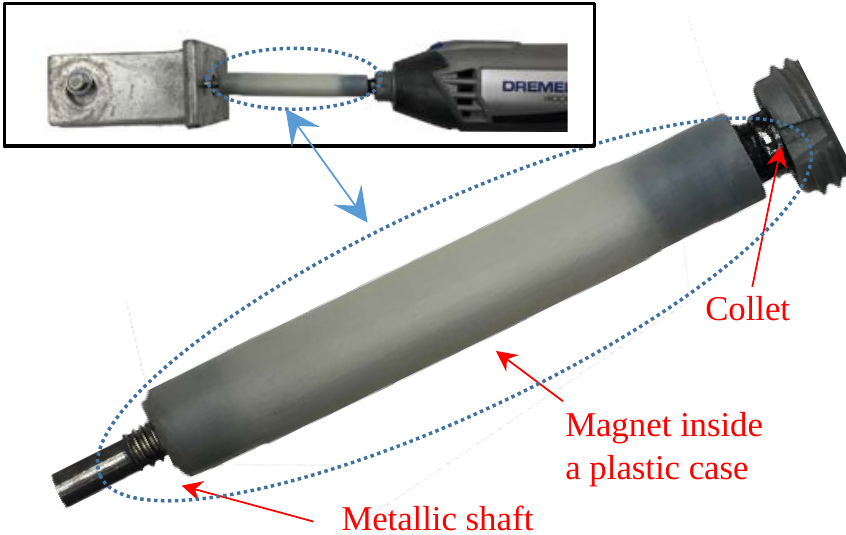}\label{fig:fig4(bc)}
    }
    
            \subfigure[]
    {
        \includegraphics[clip, trim=2.5cm 23.5cm 12.5cm 2.7cm,width=0.9\columnwidth]{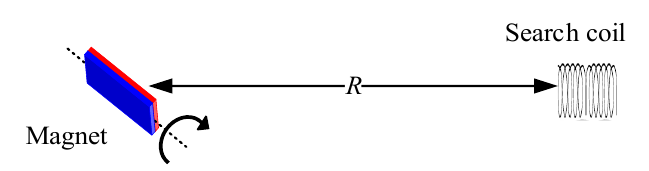}
        \label{fig:fig4(c)}
    }
  \caption{\textit{a}) Photograph of the prototyped transmitter  \textit{b}) photograph of the permanent magnet plastic case. The inset shows the photograph of magnet and Dremel 4000 rotary tool \textit{c}) Maxwell model of the magnet rotation setup.}
\end{figure}

\vspace{-1em}
Assessing the performance of the prototyped transmitter (\textcolor{cyan}{Fig.~\ref{fig:fig4(a))}}) is a significant challenge due to the lack of a reliable and calibrated magnetometer. As a result, we use the magnetic field of a rotating permanent magnet as a reference. We also used a low-noise audio amplifier connected to an air-core search coil as a receiver. Besides, the magnet used in the transmitter and the one used as the rotating magnet are identical. If the measurement setup is the same (the transmitter replaces the rotating magnet while the relative location to the search coil is the same), we can assess our transmitter accurately. \textcolor{cyan}{Figure~\ref{fig:fig4(bc)}} shows the permanent magnet plastic case, which connects to a Dremel 4000 rotary tool (35000 rpm) through its main shaft (see the inset of \textcolor{cyan}{Fig.~\ref{fig:fig4(bc)}}). A metal shaft is in place to secure the other end of the plastic enclosure to a solid fixture when it rotates. We were able to rotate the magnet up to 25800 rpm (equivalent to 430Hz). \textcolor{cyan}{Figure~\ref{fig:fig4(c)}} shows the rotating magnet and the search coil with the distance $R = 1.2\, m$.

% \begin{tikzpicture}% based on https://tex.stackexchange.com/a/9561/ (Caramdir's fantastic answer to another question)
%         \node (tiger) [anchor=south west, inner sep=0pt] {\includegraphics[width=\textwidth]{tiger}};
%         \begin{scope}[x={(tiger.south east)},y={(tiger.north west)}]
%           \foreach \i/\j in {{(0.23,1.05)/(0.25,-.1)},{(0.54,1.1)/(0.5,-.15)},{(0.76,1.05)/(0.8,-0.1)}}
%             \draw [red, thick] \i -- \j;
%         \end{scope}\textbf{}
    %  \end{tikzpicture}
% \begin{figure}
%     \centering
%     \begin{tikzpicture}
%         \node[anchor=south west,inner sep=0] (image) at (0,0) {\includegraphics[width=\columnwidth]{mushroom}};
%         \begin{scope}[x={(image.south east)},y={(image.north west)}]
%             \draw[help lines,xstep=.1,ystep=.1] (0,0) grid (1,1);
%             \foreach \x in {0,1,...,9} { \node [anchor=north] at (\x/10,0) {0.\x}; }
%             \foreach \y in {0,1,...,9} { \node [anchor=east] at (0,\y/10) {0.\y}; }
%             \node[anchor=south west,inner sep=0] (image) at (0.5,0.7) {\includegraphics[width=0.1\columnwidth]{tux}};
%         \end{scope}
%     \end{tikzpicture}
%     \caption{Find that penguin!}
% \end{figure}

\begin{figure}
    \centering
    \subfigure[]
    {
        \includegraphics[clip, trim=2.5cm 21.3cm 11.8cm 2.5cm,width=0.9\columnwidth]{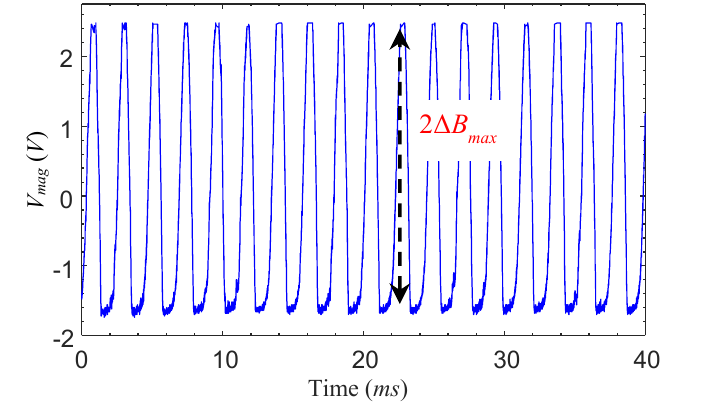}
        \label{fig:fig5(a)}
    }
    \\
    \subfigure[]
    {
        \includegraphics[clip, trim=2.5cm 21cm 11.5cm 2.5cm,width=0.9\columnwidth]{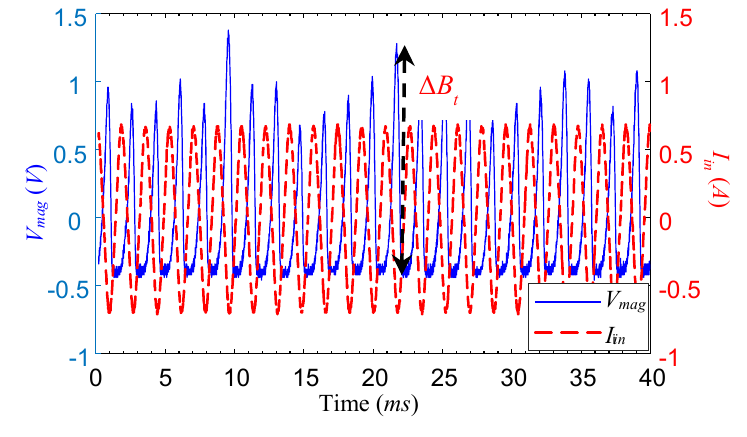}
        \label{fig:fig5(b)}
    }
\caption{Measured magnetic flux of \textit{a}) rotating permanent magnet and \textit{b}) proposed transmitter.}
\label{fig:fig5} 
\end{figure}

\textcolor{cyan}{Figure~\ref{fig:fig5(a)}} displays the measured signal at the output of the low-noise audio amplifier connected to the search coil as the rotary tool rotates the magnet. The distorted waveform is due to the non-linearity of the detection circuitry (audio amplifier). Next, we remove the rotary system and replace it with the proposed transmitter. We used a signal generator to feed the proposed transmitter via a buffer amplifier with a 430 Hz sinusoidal waveform. In addition to the voltage waveform on the audio amplifier output, \textcolor{cyan}{Fig.~\ref{fig:fig5(b)}} displays the input current waveform. Comparing the two voltage waveforms in \textcolor{cyan}{Fig.~\ref{fig:fig5}} is reasonable by maintaining the same method of receiving and measuring for both cases. Notice that the rotating magnet switches its field polarity per half a cycle (swinging between $+B(R)$ and $-B(R)$ or $2\Delta B_{max}$) while the proposed transmitter can open and close the entire magnetic flux of the magnet (swinging between 0 and $+B(R)$ or $\Delta B_{t}$) at its peak. Therefore the rotating magnet produces twice as much a time-varying magnetic flux as our proposed transmitter produces at its ideal performance. Besides, the magnetic flux maximum $\Delta B_{max}$ is equal to its static value for a given permanent magnet, due to the low frequency (quasi-static). Simply, a rotating magnet's time-variant magnetic flux is equal to $B_{max}\, cos \omega t $. From now on, we compare the transmitter's measured time-variant flux with the permanent magnet's static flux at the same point, and we call it modulation depth.
\begin{equation}
Modulation\;depth = \frac{\Delta B_{t}}{\Delta B_{max}} \times 100 \quad \left (\% \right)
\label{eq:four}
\end{equation}

\begin{figure}
    \centering
        \includegraphics[clip, trim=5cm 20.3cm 8.5cm 1.5cm,width=0.9\columnwidth]{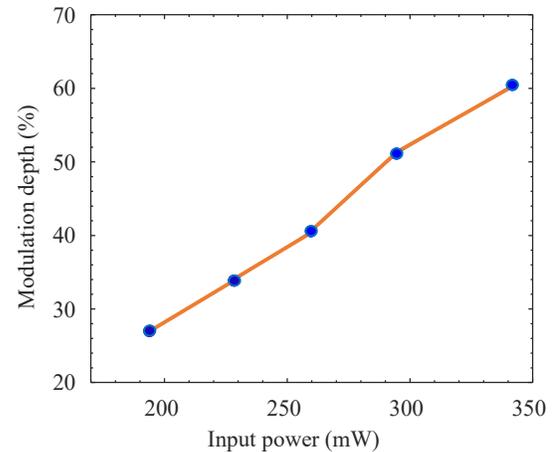}
\caption{The modulation depth as a function of the input power of the transmitter.}
\label{fig:fig6} 
\end{figure}

Measuring the total power required to generate a time-varying magnetic flux at a given distance is a crucial factor in evaluating the transmitter's performance. Based on the measured signal shown in \textcolor{cyan}{Fig.~\ref{fig:fig5(b)}}, the sinusoidal voltage applied to the control coil is 0.95 V, and the current is 0.6 A, which results in an average power of 0.285 W, while the rotary device needs 60 W to rotate the magnet. The modulation depth of the proposed transmitter and the rotary device can be compared with the measured input power in mind. The measured flux, shown in \textcolor{cyan}{Fig.~\ref{fig:fig5}}, used to calculate the modulation depth of 51\%. Note that the maximum modulation depth for the transmitter is 100\%, while the magnet's modulation depth is 200\%. \textcolor{cyan}{Figure~\ref{fig:fig6}} also shows the measured modulation depth of the transmitter versus the input power.

\begin{figure}
    \centering
        \includegraphics[clip, trim=2.5cm 22.2cm 12.5cm 2.5cm,width=0.9\columnwidth]{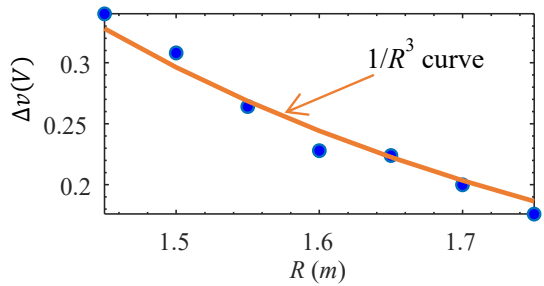}
\caption{The measured field versus range. The data points show each individual measurement, and the line is the result of curve fitting.}
\label{fig:fig7} 
\end{figure}

One approach to verifying the measurement method is to measure the magnitude of the magnetic flux at various distances for a given sinusoidal drive current. \textcolor{cyan}{Figure~\ref{fig:fig7}} shows the output voltage of the receiver vs. {\it R}. The magnetic flux (which is linearly proportional to the output voltage) decays by $1/ R^3$ as expected. Besides, this figure provides a guideline for estimating the magnitude of the time-variant magnetic flux at any distance where measured/simulated data at least at one point in the same direction is available. The theoretical equation \cite{Manteghi17} was used to find the magnetic flux for the rotating magnet and then to determine the coefficient required to convert the obtained voltage to the magnetic flux.

The transmitter will, therefore, generate $0.17\, \mu T$ at 1.2 m. In the same way, the $1/ R^3$ decay of the magnetic field of the antenna allows extrapolating the field at a distance of 1 km, although the magnetic flux of 1 km is too low to measure with our magnetometer. It is estimated that the magnetic flux will be 0.294 fT at 1 km. This study is conducted to determine the magnet volume needed to achieve a field strength of 1 fT at 1 km. The results show that 1 fT can be accomplished at 1 km with a permanent magnet volume of $10\,cm^3$ with a power consumption of less than 0.5 W. Also, the proposed antenna is compared with other current designs in \textcolor{cyan}{Table~\ref{tab:table1}}. The magnetic field generated by volume ($\Delta B/V$) for different designs shows that the rotary magnet systems produce the maximum field with a range of approximately $0.2\, fT/cm^3$. However, this technique has its limitations. The multiferroic transmitter, which generates a magnetic field of approximately $0.013\, fT/cm^3$, is also far from competing with the rotating magnet. The proposed transmitter in this paper can generate a $0.1\, fT/cm^3$ magnetic field, making it a feasible candidate to compete with the rotating magnet.

In terms of bandwidth and data rate, the proposed transmitter does not comply with the fundamental antenna limits. The conventional antenna design approaches depend on the practical and useful Linear Time-Invariant (LTI) systems. For example, a lossless tuned electrically small antenna (ESA) at resonance can be treated as a second-order resonator, where the stored electrical/magnetic energy in its reactive zone exchanges the stored magnetic/electric energy in the reactive lumped element of the antenna’s matching circuit. For example, a 1-meter lossless resonant antenna at 1 kHz ($\lambda = 300 m$) has a minimum Q of $10^{14}$ (bandwidth of $10^{-11}\, Hz$). However, bandwidth can be increased by sacrificing the antenna efficiency that can be achieved only on the receive side, but not on the transmitter. However, it has been shown that the fundamental limits of the antennas do not bound the non-linear and/or time-variant (non-LTI) antennas \cite{Manteghi19}. 

For example, a time-variant field can be created while avoiding resonance, if the stored energy in an antenna's reactive near-zone does not transform into another type of energy every half a cycle (first-order system), and time variation is realized by changing the location where the energy is stored. Therefore, the time-variant basis of the proposed structure gives rise to a parametric or non-LTI system that allows us to change the data transfer rate, independently from the antenna quality factor. As a consequence, this non-LTI system results in higher data rates being feasible. Moreover, it has shown that the stored energy frequency can be quickly shifted (FSK) without breaching the fundamental limits \cite{Salehi13}. Therefore, the frequency of the field modulation in the proposed transmitter can be changed from a few hundred hertz to tens of kilohertz without any restriction. Besides, any type of modulation, such as frequency or amplitude modulation, can be applied to the proposed transmitter.
\vspace{-1em}
\section{\label{sec:simulation} Simulation Results}
\begin{figure}
    \centering
    \subfigure[]
    {
        \includegraphics[clip, trim=2.5cm 21.2cm 12.5cm 2.3cm,width=0.9\columnwidth]{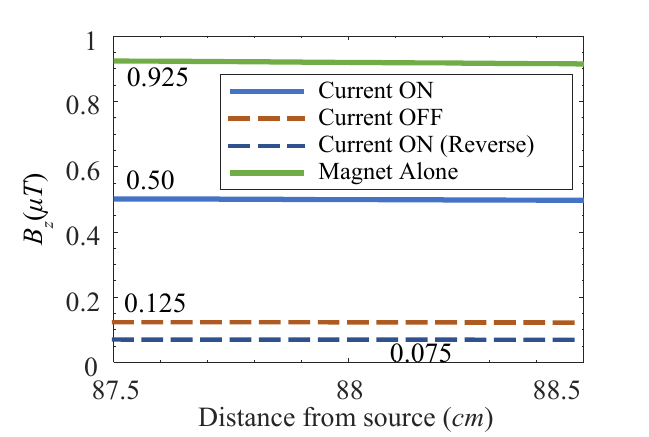}
        \label{fig:fig8(a)}
    }
    \\
    \subfigure[]
    {
        \includegraphics[clip, trim=2.5cm 22.7cm 12.8cm 2.5cm,width=0.89\columnwidth]{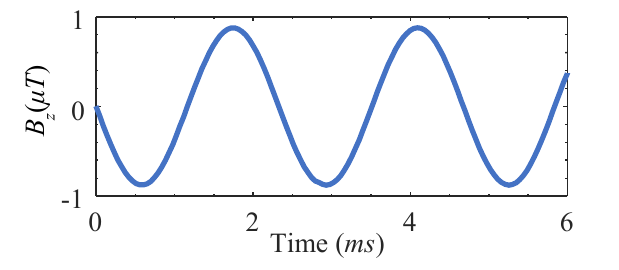}
        \label{fig:fig8(b)}
    }
    \subfigure[]
    {
        \includegraphics[clip, trim=2.7cm 22.7cm 12.9cm 2.3cm,width=0.85\columnwidth]{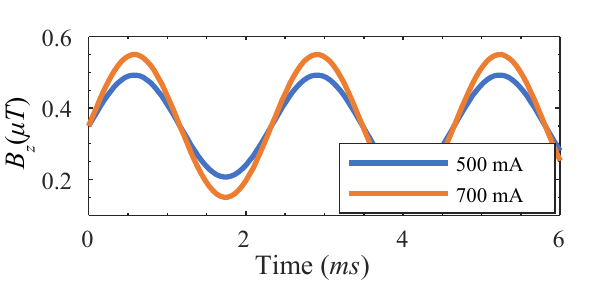}
        \label{fig:fig8(c)}
    }
\caption{Simulation results; \textit{a}) Magnitude of the magnetic flux at 0.88 m away from the magnet in the magnetostatic solver, \textit{b}) time domain solution of the rotating magnet, and \textit{c}) time domain solution of the transmitter in the transient solver, for two different control currents.}
\label{fig:fig8} 
\end{figure}

\vspace{-1em}
We conduct further analysis in the simulation domain after verifying the transmitter's functionality in the measurement domain. We used magnetostatic simulation in the software package, ANSYS Maxwell, to achieve that objective. In this analysis, four different cases have been simulated: 1- an isolated permanent magnet, 2- an open mode transmitter (current ON), 3- a closed mode transmitter (current OFF), and 4- a deep closed mode transmitter (reverse current ON). One can use the case 1 magnetic flux to examine the effects of the electric current and the magnetic film thickness on magnetic flux in case 2 and case 3. Also, the modulation depth is determined by subtracting from case 2 the magnetic flux in case 3 or 4 and dividing the result by case 1 magnetic flux. The simulation results for different cases at $R = 0.88\,m$ are shown in \textcolor{cyan}{Fig.~\ref{fig:fig8}}.

\textcolor{cyan}{Figure~\ref{fig:fig8(a)}} shows that case 2 (open-mode transmitter) generates 54\% of the flux from an isolated magnet (case 1). This value is essential as we determine the size of the magnet required for a given application. Besides, the modulation depth for case 3 and case 4 is 41\% and 46\%, respectively. Although we used an approximate B-H curve for the Metglas film in the simulation domain, the results are in good agreement with the measured results (51\% modulation depth). Note that the drive current is a balanced sinusoidal in our measurement setup (plus and minus currents); therefore, we compare the measured results with modulation depth in case 4 as 46\%. Next, we analyze the time-domain behavior of the rotating magnet and the proposed transmitter using a transient analysis by ANSYS Maxwell. \textcolor{cyan}{Figure~\ref{fig:fig8(b)}} shows the magnetic flux of the rotating magnet at {\it R} = 0.88 m. As we expected for a quasi-static case, the maximum value of the flux is equal to the magnetic flux of the static magnet at the same distance {\it R} = 0.88 m. The same behavior is observed for the proposed transmitter for two different drive currents.

\begin{figure}
    \centering
        \includegraphics[clip, trim=2.5cm 20.3cm 12.8cm 2.3cm,width=0.9\columnwidth]{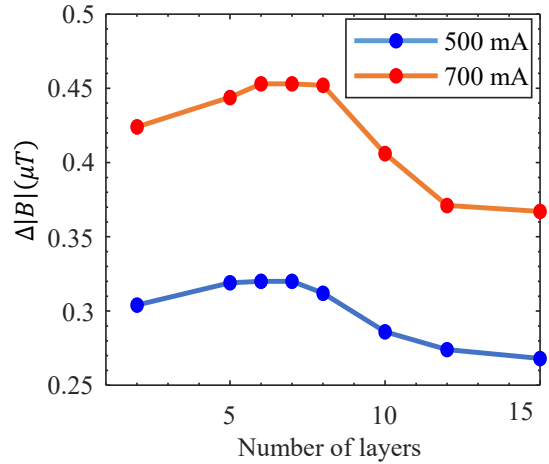}
\caption{Effect of the number of layers in modulation depth, when 0.5 A and 0.7 A, are applied as input control current.}
\label{fig:fig9} 
\end{figure}

We have also analyzed the effect of the magnetic film thickness on the transmitter performance. The Metglas film available comes in a roll, with a thickness of 10 mil (0.0254 mm). The thickness of the magnetic film can, therefore, vary from one layer to an integer number of layers $n \times 10\, mil$. \textcolor{cyan}{Figure~\ref{fig:fig9}} shows the simulation results for a variety of Metglas layers used in the magnetic film for two different drive currents. The optimal number of layers for drive current of 500 mA and 700 mA is 7 and 8, respectively. Therefore, to build the magnetic film, one has to know the drive current in addition to the magnetic material's B-H curve. 
\vspace{-1em}
\section{\label{sec:conclusion} Conclusion}
\vspace{-1em}
A new method for generating electromagnetic waves using the permanent magnet's static magnetic flux has been introduced. By using reluctance modulation, the direction of the magnetic flux and the location of the stored magnetic energy have been changed to create a time-variant field. A method for evaluating a ULF transmitter's performance has also been implemented and used to assess the proposed transmitter. It has been shown that the prototype transmitter produces a time-variant field with a modulation depth of 50 percent. While we have not tried to minimize the size and weight of the transmitter, it has realistic dimensions and weight. We also analyzed the power consumption of the transmitter and the calculated results. The calculations show that we can generate 1 fT of time-variant magnetic flux at 1 km using a magnet volume of $10\,cm^3$.  

The data that support the findings of this study are available from the corresponding author upon reasonable request.
\vspace{-1em}

% \begin{acknowledgments}
% We thank Prof. Khai Ngo for his support and valuable advice on this work.
% \end{acknowledgments}
\vspace{-1em}
% \nocite{*}
\section*{References}
\vspace{-1em}
\bibliography{aipsamp}% Produces the bibliography via BibTeX.

\end{document}